\documentclass[%
 reprint, prl,
 amsmath,amssymb,
 aps,
]{revtex4-2}

\usepackage{graphicx}
\usepackage{dcolumn}
\usepackage{bm}
\usepackage{xcolor}

\begin{document}

\preprint{APS/123-QED}

\title{The role of spatiotemporal couplings in harmonic vortex generation}

\author{C. Granados$^{1}$}
\email{cagrabu@eitech.edu.cn}
\author{Bikash Kumar Das$^{2,3,4}$}
\author{Marcelo F. Ciappina$^{2,3,4}$}
\author{Wenlong Gao$^{1}$}

\affiliation{$^{1}$Eastern Institute of Technology, 315200, Ningbo, China}
\affiliation{$^{2}$ Department of Physics, Guangdong Technion-Israel Institute of Technology, 241 Daxue Rd.,515063,  Shantou, China}
\affiliation{$^{3}$Guangdong Provincial Key Laboratory of Materials and Technologies for Energy Conversion, 241 Daxue Rd.,515063,  Shantou, China}
\affiliation{$^{4}$ Department of Physics, Technion-Israel Institute of Technology, 32000, Haifa, Israel}

\date{\today}

\begin{abstract}
We explore the impact of spatiotemporal couplings (STCs) on high-order harmonic generation (HHG) driven by spatiotemporal vortex beams. Our investigation demonstrates how STCs shape key properties of the generated harmonic beams, including their intensity distribution and different chirps. By analyzing these chirps, we establish a clear connection between STCs and the observed harmonic structures. Furthermore, we examine the HHG process in both the near- and far-fields, identifying the conditions under which these perspectives align and provide consistent results. By clarifying the role of spatiotemporal vortex beams in HHG, this work contributes to a broader understanding of the interplay between spatiotemporal effects and harmonic generation, while offering a framework to merge differing interpretations in the literature.

\end{abstract}

\maketitle


\textit{Introduction:} The development of structured light, characterized by its ability to carry orbital angular momentum (OAM), introduced the possibility of extending light-matter interactions into the topological realm. Whether utilized as a light source \cite{TunableVort} or directly for spectroscopy \cite{ChiralMol}, structured light serves as a powerful tool for investigating how the topology of an external electromagnetic field can probe and manipulate the structure of atoms, molecules, and solids \cite{ChiralMol, NonLinHeli, SolidFoil, FerroMagneticHHG, CrystallineSolids, SBE_Solids}.
The capability to shape light in both space and time has become a routine practice in specialized laboratories \cite{TSTOV, STOV_NatPhot}, where generating various vortex beams-from simple Laguerre-Gauss (LG) or Bessel-Gauss (BG) beams to more complex toroidal or conch-shaped structures \cite{TSTOV, NSBessel, STOV_NatPhot, Conch}—is now achievable. These offer a diverse range of spatial and temporal structures and characteristics that remain to be fully explored in light-matter interactions. More importantly, using structured light as the driving field facilitates the generation of structured attosecond pulses in the extreme ultra violet (XUV) regime~\cite{VortexXUV, AttoVortex, AttoVortTrains}. Spatial (or longitudinal) vortex beams are characterized by an OAM projected along the propagation direction, featuring a singularity at the center of the beam. Around the singularity the wavefront of the light beam twists. The number of these twists per unit wavelength defines the topological charge (TC) of the beam. This TC dictates the helical phase of the vortex beam, represented as $\exp(-i l \phi)$. The phase $\phi$ contributes to the topology of the different beams by the coupling of the transverse spatial coordinates $x$ and $y$ as follows: $\phi = \arctan(y/x)$. In the case of spatiotemporal (or transverse) optical vortex (STOV) beams \cite{STOV2012}, it is necessary to introduce the  moving frame (or local) time $\tau = c(t-\delta t)$ in addition to the spatial coordinates \cite{STOV2012,FirstSTOV,PropSTOV}. Unlike spatial vortex beams, STOV beams represent a spatiotemporal structure that carries OAM in a direction transverse to the beam's propagation. An important question arises here: What is the effect of using STOV beams in a nonlinear process such as high-order harmonic generation (HHG)?. We can anticipate an important characteristics of HHG process driven by an STOV beam, and it is the coupling between the spatial and temporal coordinates through the helical phase. This leads to an electromagnetic field with non-separable space-time coordinates resulting in spatiotemporal couplings (STCs). For Gaussian ultrashort pulses, STCs are often overlooked. Despite this, STCs can be synthesized through various methods \cite{ST_Control1, ST_Control2, ST_Control3} for the generation of isolated attosecond pulses \cite{LHExp, LHExp2, LHExp3}. For example, a slight misalignment in a laser's compressor, as shown in Ref. \cite{LHeffect}, causes STCs to appear in the harmonic spectrum as spatially separated attosecond pulses. This effect is known as the lighthouse effect \cite{LHeffect} and leverages on the wavefront rotation \cite{trevino2}. Additionally, first-order spatiotemporal couplings (STCss) are well-characterized in terms of Fourier transforms (FTs) and the resulting chirps~\cite{trevino,trevino2, HullierSTC, RhodesSTC}. 


The interaction of spatiotemporal vortices with matter during the HHG process has been described at the single-atom level and in its near-field using the quantum orbit (QO) version of the strong-field approximation (SFA) \cite{PKU, HHGHofnions, QO, SFA_OV}. The QO approach focuses on the microscopic temporal response of the atom to the STOV by incorporating directly the electromagnetic field in the time-dependent dipole response. In spite of that, the interpretation of the intensity distribution in the spatiospectral domain could be seen as ambiguous and may benefit from further clarification.
Complementary to the near-field analysis, the HHG process can also be described in the far-field using the thin-slab model (TSM) \cite{VortexXUV}. Notice that in any description of the HHG process, the experimental measurements are performed in the far-field. Nevertheless, near- and far-field descriptions should lead to similar conclusions. The TSM model describes the harmonic generation process by approximating the harmonic field by the fundamental field to a power given by the a intensity scaling law. The scaling law can be derived from either the SFA or the time-dependent Schrödinger equation. Using the TSM, the macroscopic response (far-field) is computed through the Fraunhofer far-field integral \cite{AttoVortex}. Both the QO and TSM approaches neglect the STOV beam's spatial structure and its coupling to the atom, relying on the dipole approximation, which is justified as the beam is much larger than the electron excursion.

This letter investigates STCs in HHG, reinterpreting recent near-field results for STOV-driven HHG \cite{PKU, AttoScheme}. We show that harmonic vortices’ chirp and intensity distribution arise from Fourier transform properties rather than STOV beam interference. Additionally, we highlight the importance of considering the far-field counterpart to avoid misinterpreting STCs in HHG. Our work broadens the understanding of STCs in STOV-driven HHG and opens new avenues for engineering light topologies and attosecond pulses.


\textit{Spatiotemporal couplings:} We will restrict to the spatiospectral and $t-k_x$ domains, since it is convenient for the description of the HHG process. As noted in Refs.\cite{trevino, LHeffect, STOV_NatPhot}, STC defines a new type of laser field whose spatial and temporal properties are not separable, i.e., $E(x,t) \neq E(x)E(t)$. 
STCs appear naturally in the STOV electromagnetic field~\cite{OSTV} and are non-separable, as space and time are coupled by the helical phase. This is evident in the general function describing a STOV beam:
\begin{eqnarray}
	E_{STOV}(x,y,z,t) &=& A(x,y,z,t) \exp{\left( - i l \phi \right)} \nonumber \\ 
	&\times& \exp{(-i k z - i\omega_0 t)}, \label{stbeam}
\end{eqnarray}
here, $r^2=x^2+y^2+z^2+\tau^2$ is the spatiotemporal coordinate and $A(x,y,z,t)$ is the envelope function of the STOV. The TC is represented by $l$, $z$ is the propagation direction and $\omega_0$ the fundamental laser frequency. We start examining the role of the STCs in HHG by setting it to zero, i.e. using $l=0$. 
We will use a Gaussian envelope here unless stated otherwise, given by $ A(x,y,z,t) = E_0(t^2/w_t^2+x^2/w_x^2)^{|l|/2}\exp{(-x^2/w_x^2-\tau^2/w_t^2)}$, where $w_t$ and $w_x$ are the temporal and spatial width, respectively. The STOV with $l=0$ is called a spatiotemporal wave-packet (STWP) with an amplitude term equal to $E_0$. In this case, Eq.~(\ref{stbeam}) takes the form of a separable Gaussian pulse, i. e., $E_{STOV}(x,y,z,t) = E_{STOV}(x,y,z)E(t)$. Thus, in any Fourier domain, the STWP behaves as a Gaussian distribution, either elongated or compressed in reciprocal space, depending on the spatial and temporal widths.
However, if $l\neq 0$, Eq.~(\ref{stbeam}) describes an STOV field with STCs, which appear in different Fourier domains as distorted fields or chirps. The spatiotemporal Fourier transform (FT) of the STOV beam corresponds to Hermite polynomials, as demonstrated in Ref.~\cite{PorrasSTOV}.
\begin{figure}[h!]
\centering 
\includegraphics[width=1.0\linewidth]{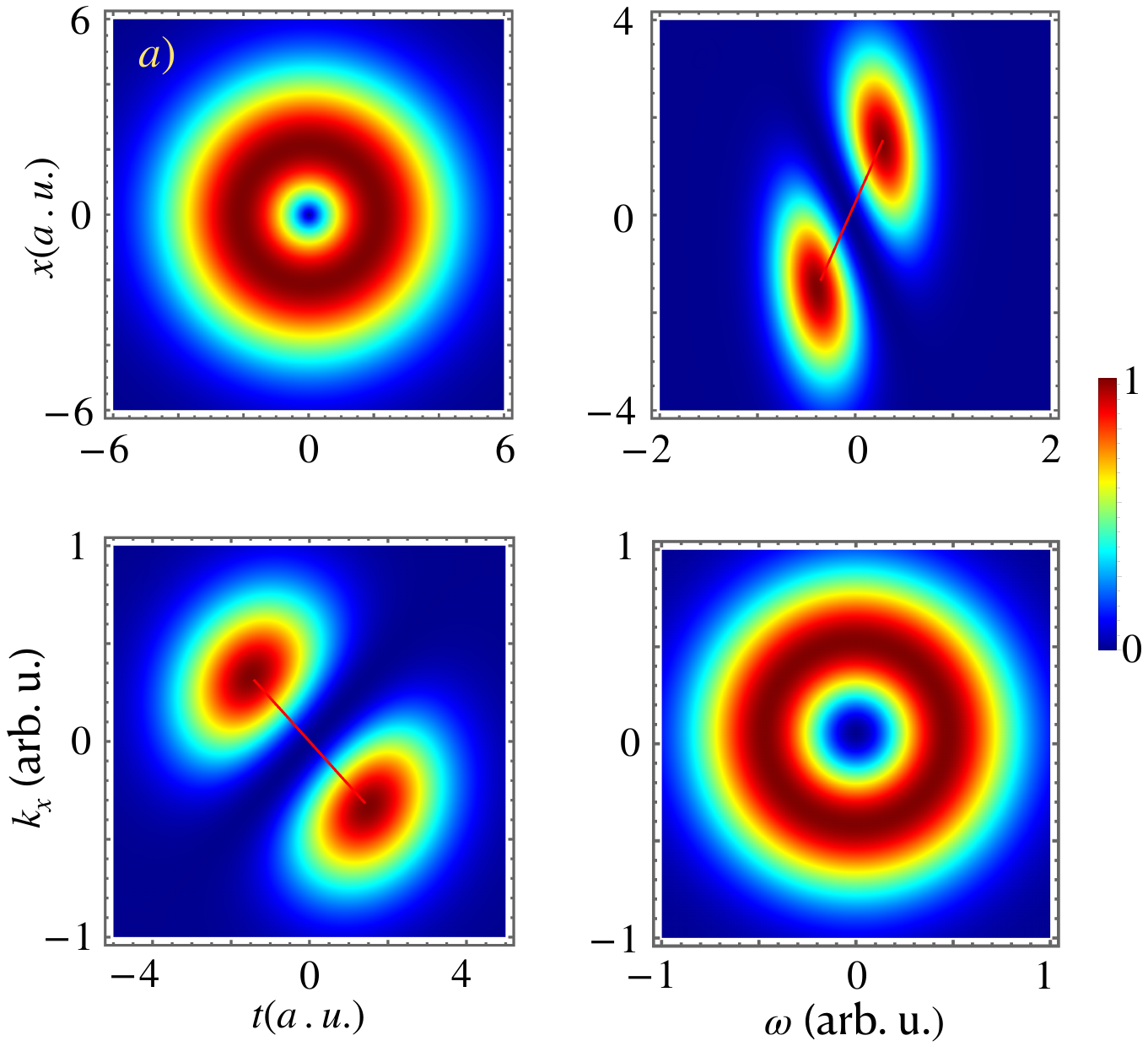}
\caption{Normalized spatiotemporal vortex beam in different Fourier domains. (a) the spatiotemporal domain; (b) the time-momentum domain; (c) the spatiospectral domain and (d) the momentum-spectral domain. The red lines in (b) and (c) represent the chirp function given by $x_{FT}(\omega)$ and $k_x(t) = -2ct/w_\xi w_x$. }
\label{Fig1}
\end{figure}
\vspace{0cm}
In Fig.~\ref{Fig1}, we present the numerical FT of Eq.~(\ref{stbeam}) in different Fourier domains, starting from the spatiotemporal domain. For simplicity, we set $y=0$. It is evident that when the STCs are nonzero, the beam intensity distribution exhibits spatial, spectral, and angular chirps across different domains.  Additionally, the TC of the beam is reflected in the $t-k$ and $x-\omega$ domains as minima between maximum intensity lobes. In the case shown in Fig.~\ref{Fig1}, one can infer a TC value of $l=1$. For plotting Fig.~\ref{Fig1}, we used a wavelength of 800~nm and set the temporal and spatial widths in Eq.~(\ref{stbeam}) to 3~a.~u. and $411$~a.~u., respectively.   Our analysis clearly shows that the lobed structure of the STOV beam in the $x-\omega$ and $k_x-t$ domains appears only when STCs are present in the fundamental beam, with the number of minima indicating the TC of the beam. Furthermore, we demonstrate that STCs arise from the helical phase in STOV beams and can be controlled either by the TC or by adjusting the spatial and temporal widths. In conclusion, when STCs are absent ($l=0$), the intensity distribution remains Gaussian in all Fourier domains.  

\begin{figure}[h!]
\centering 
\includegraphics[width=0.7\linewidth]{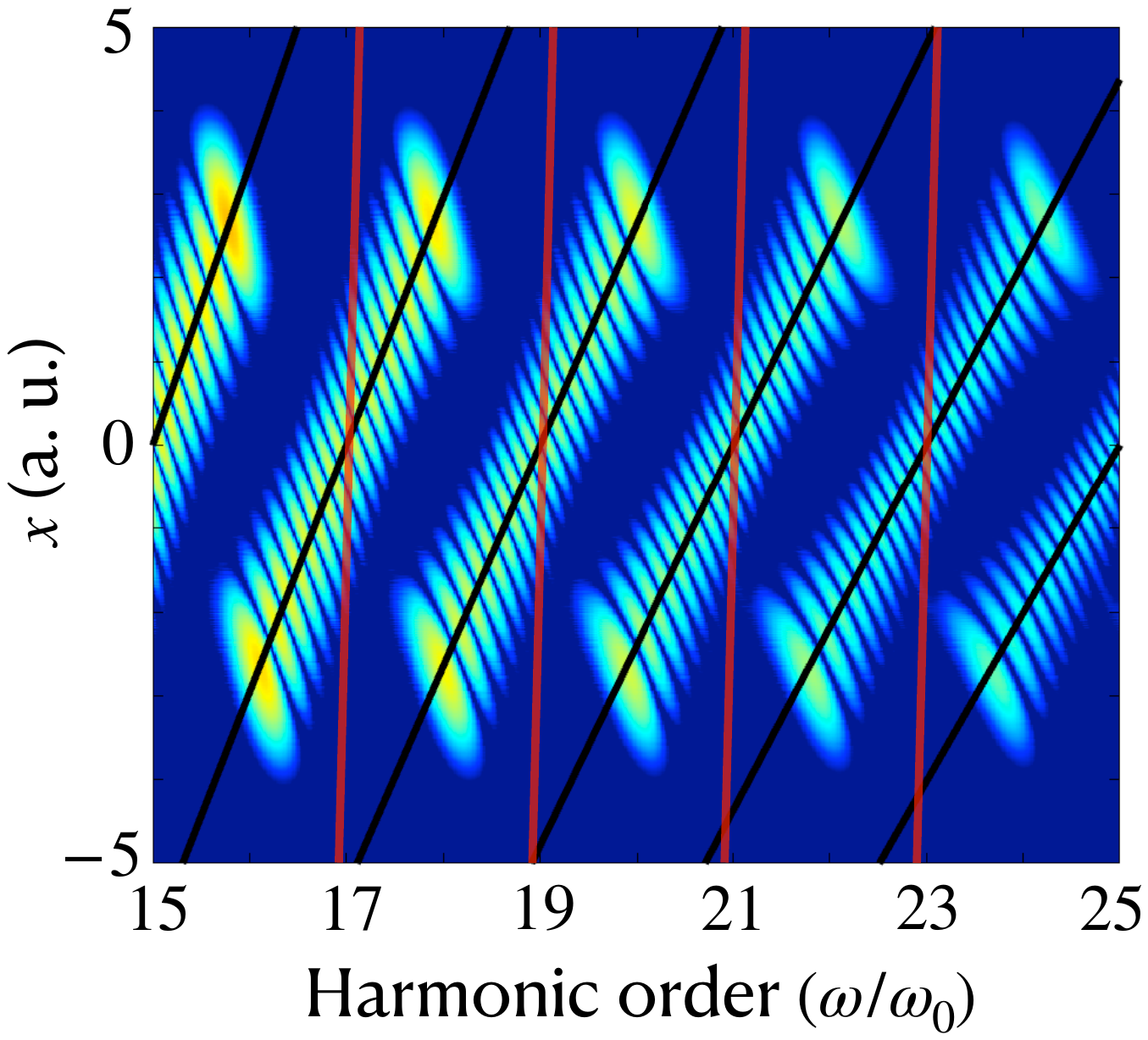}
\caption{Harmonic vortices in the near-field. The driven field is shown in Eq.~(\ref{stbeam}). Here, $l=1$, $\lambda=400$~nm and $E_0=0.12$~a. u. The red and black lines represent the chirp functions $x_{FT}(\omega)$ and $x_{QO}(\omega,q)$, respectively.}
\label{Fig3}
\end{figure}

\textit{Near-field calculations:} What is the effect of STCs on the HHG process? To answer this, we examine the near-~\cite{PKU} and far-field~\cite{VortexXUV}. We begin with the near-field, which can be obtained using either the QO version of the SFA~\cite{QO} or the TSM. Figure~\ref{Fig3} presents results for $l=1$ using the QO approach, reproducing the findings of Ref.~\cite{PKU}. To account for the spatial chirp in the near-field harmonics, we analyze the FT of the fundamental field, Eq.~(\ref{stbeam}). For a TC value $l$, its FT is given by~\cite{PorrasSTOV}:  
\begin{eqnarray}
    \widetilde{E}(x,\omega) &\propto& \frac{E_0}{2^l} \exp\left(-\frac{w_\xi^2}{2c^2}(\omega-\omega_0)^2\right)\exp{\left( - \frac{x^2}{w_x^2}\right)}\nonumber \\ 
    &\times&H_l\left(\frac{w_\xi}{2c}(\omega-\omega_0) - \frac{x}{w_x} \right), \label{FT_STOV}
\end{eqnarray}
where $H_l(...)$ is the Hermite polynomial of degree $l$, representing a tilted intensity distribution with lobes. The number of minima between lobes corresponds to the TC value. To derive the chirp function describing the harmonic slope, we set the argument of $H_l$ in Eq.~(\ref{FT_STOV}) to zero and solve for $x$. For $l=1$, this yields:  
$x_{FT}(\omega) = \frac{w_\xi w_x (\omega-\omega_0)}{2c} \times \frac{\omega_0}{\lambda}$. However, reproducing the correct slope requires incorporating the harmonic order, leading to a spatial chirp of the form $x_{QO}(\omega,q) = \frac{w_\xi w_x (\omega-\omega_0)}{2qc} \times \frac{\omega_0}{\lambda}$.  
In Fig.~\ref{Fig3}, the red and black lines represent $x(\omega)$ and $x(\omega,q)$, respectively. Interestingly, the TC has little effect on the slope but primarily shifts the intersection with the $x$-axis. As shown in Fig.~\ref{Fig3}, the agreement of $x(\omega,q)$ with the harmonic field tilt confirms that spatial chirp originates from the STCs inherent in the driving spatiotemporal field. The dependence of spatial chirp on harmonic order requires further verification using the TSM in both near- and far-field analyses.

\begin{figure}[h!]
\centering 
\includegraphics[width=1.0\linewidth]{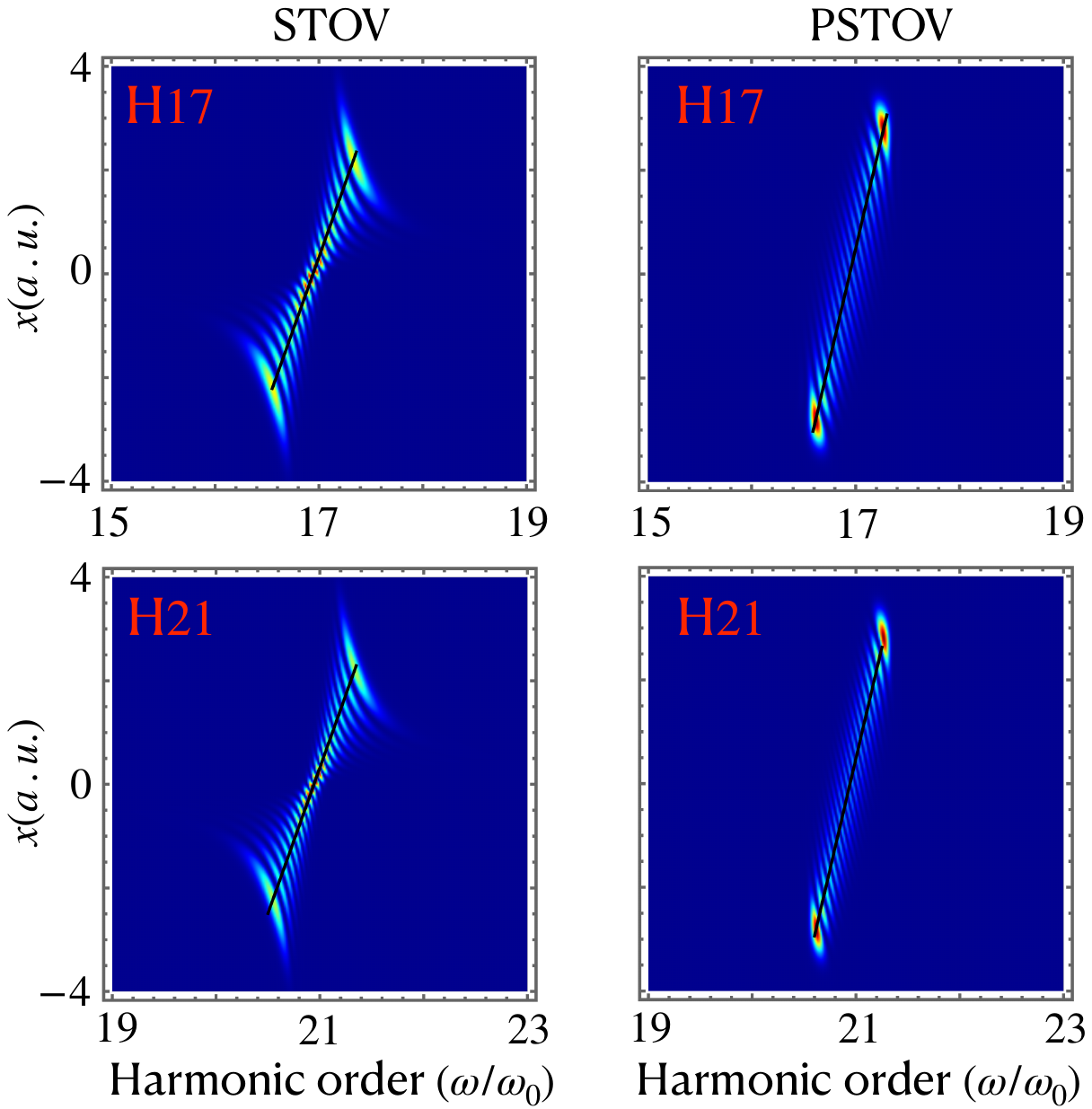}
\caption{Harmonic vortices in the near-field. Left column:  harmonic orders 17$^{\text{th}}$ (upper panel) and 21$^{\text{st}}$ (lower panel) generated with a driving STOV beam. Right column: same harmonics for a driving PSTOV beam. The black lines represent the chirp functions $x_{TSM}(\omega)$ for each case (see the text). } 
\label{FF}
\end{figure}
The TSM describes the harmonic amplitude and phase in the near-field by a structured electromagnetic field as: $E^{near}_{STOV}(x,t)=\left(A(x,y,z,t)\right)^p\exp(-iql\Phi)$, with all the phase terms condensed in $\Phi$. To compare the QO and TSM results, we calculated the near-field harmonics in the spatiospectral domain $\mathcal{F}(E^{nf}_{STOV}(x,t)) =\widetilde{E}^{nf}_{STOV}(x,\omega) $: 
\begin{eqnarray}
    \left|\widetilde{E}^{near}_{STOV}(x,\omega)\right|^2 &\propto& \Bigg|\int dt \left(A(x,y,z,t)\right)^p \exp{\left( - i q l \phi \right)}\nonumber \\
    &\times& \exp{\left(-i q t(\omega+\omega_0) \right)}\Bigg|^2.
\end{eqnarray}
Notice the factor $q$ in the exponential term. Here, $E_{STOV}(x,t)$ follows Eq.~(\ref{stbeam}) for $y=z=0$. The results, shown in Fig.~\ref{FF}, present harmonic orders 17$^{\text{th}}$ and 19$^{\text{th}}$ for both an STOV and a perfect STOV (PSTOV)~\cite{PSTOV}. Including PSTOV-driven harmonics demonstrates that FT analysis also applies to STCs in more complex beams.  For the calculations, we used a scaling parameter $p=3.4$ and the same temporal and spatial widths as in Fig.~\ref{Fig3}. The FT analysis for the TSM in the near-field shows that the chirp function is $x_{TSM}(\omega) = \frac{w_\xi w_x (\omega-\omega_0)}{2c}$,  
which explains the QO model results. More importantly, the harmonic chirp function follows from the FT of the fundamental field, indicating that the STCs of the driving field are imprinted identically on all harmonic vortices. Additionally, the lobed structure of the harmonic field does not result from interference within the vortex beam~\cite{PKU} but is a spectral signature of orbital angular momentum conservation, $l_q = ql$. Thus, STCs in the spatiospectral domain ($x-\omega$) manifest also in the slope of the Hermite polynomial, highlighting their role in harmonic generation.  

\textit{Far-field calculations:} 
How do spatiotemporal couplings influence the structure of harmonic vortices in the far-field? To answer this, we first clarify two key points: 1) Ref.~\cite{STOV_Exp} presented the first experimental measurement of HHG driven by an STOV beam, which corresponds to observe the process in the far-field. Thus, any theoretical description of harmonic vortex generation must be analyzed in the far-field. 2) Ref.~\cite{PorrasSTOV} provided a rigorous theoretical analysis of the fundamental field’s propagation and its FT to the spatiospectral domain. They demonstrated that an STOV beam is conjugate to a Hermite-lobed spatiospectral vortex, meaning STCs entirely dictate the beam's properties in different Fourier domains. This aligns with our near-field findings, where we showed that STCs are independent of harmonic order and govern the chirp and intensity distribution of the harmonics. While the harmonic vortices intensity distribution and chirp differs between the near- and far-field due to scaling laws and harmonic order, the far-field should exhibit similar functionality: lobed intensity distributions in the far-field arise from STCs, not from interference between multiple emission sources (different parts of the vortex beam). Likewise, the chirp in intensity distribution follows from the functionality of the FT of the fundamental field. If the multiple-source interference explanation proposed in the literature were correct, different harmonics would exhibit the same number of lobes, contradicting the well-established conservation of OAM in harmonic vortices. Additionally, using the TSM ensures that far-field results do not account for different parts of the vortex beam, as the TSM does not explicitly incorporate the vortex field in the SFA solution.  

\begin{figure}[h!]
\centering 
\includegraphics[width=1.0\linewidth]{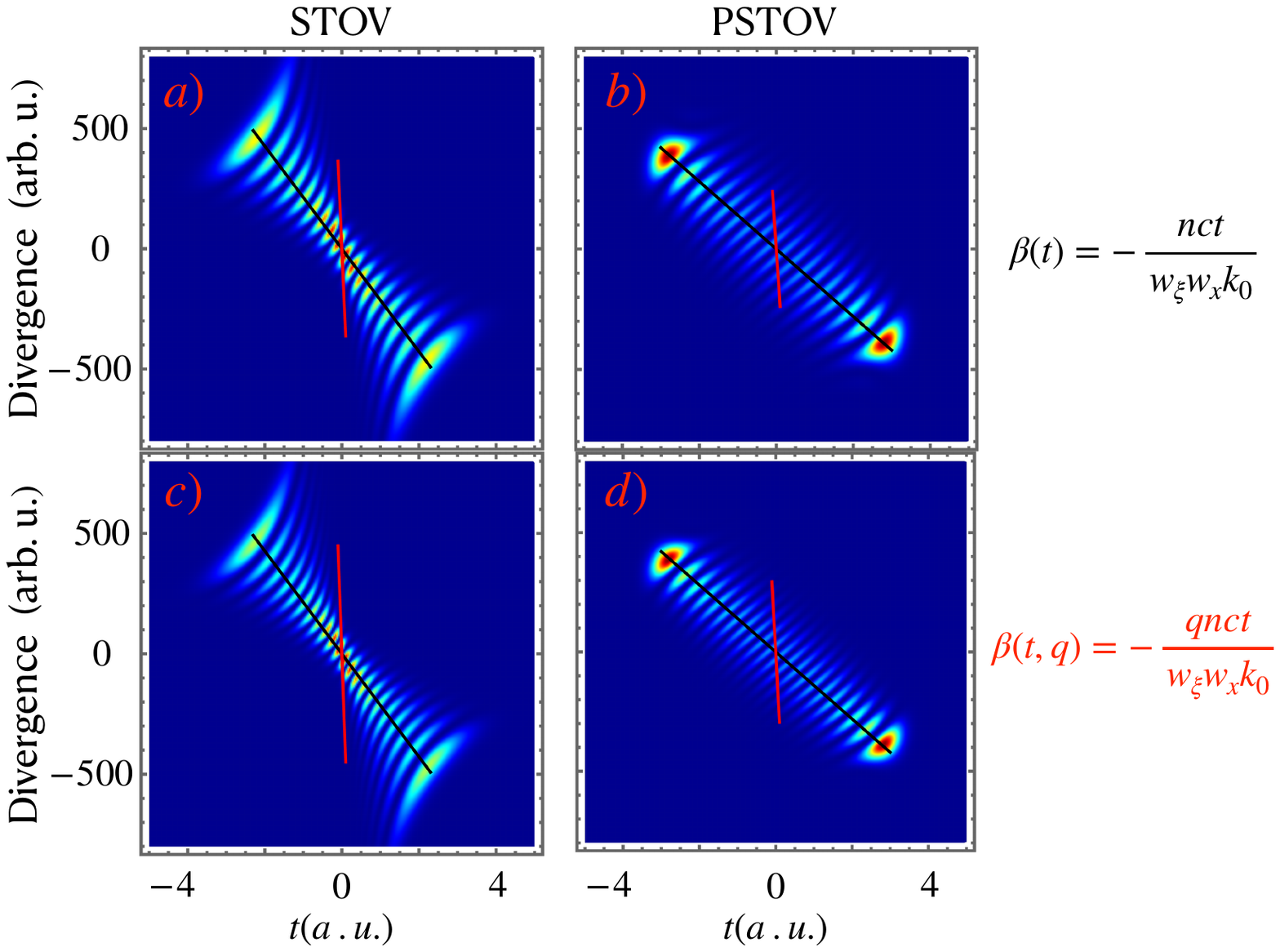}
\caption{Harmonic vortices in the far-field. (a) and (b) the 17$^{\text{th}}$ harmonic vortex generated by the STOV and PSTOV beams, respectively. (c) and (d) the 21$^{\text{st}}$ harmonic vortex. The black (red) line represents the chirp function 
$\beta(t)$ ($\beta(t,q)$) on the right-hand side. Here $n=1.6$ for the STOV and $n=1.05$ for the PSTOV.} 
\label{FF2}
\end{figure}

We can write the TSM far-field integral to calculate the harmonic field in the following way \cite{STOV_Exp}: 
\begin{eqnarray}
   \Bigg| \widetilde{E}^{far}_{STOV}(\beta_x,t)\Bigg|^2 &\propto&\Bigg| \int dx E^{near}_{STOV}(x,t) \exp{\left( - i q l \phi \right)}\nonumber \\
    &\times& \exp{\left(-i q \omega_0 t \right)}\exp{(-ik_q \beta_x x)}\Bigg|^2,\label{ffintegral}
\end{eqnarray}
where $\beta_x$ is known as the divergence, $k_q = qk_0$ and $k_0 = 2\pi/\lambda$. In Fig.~\ref{FF2}, we present the numerical solution of Eq.~(\ref{ffintegral}). The left column shows harmonic vortices driven by the STOV beam, while the right column displays those driven by the PSTOV beam. In all panels, the chirp function, shown on the right side of Fig.~\ref{FF2}, is represented by straight lines. Two key conclusions emerge from the far-field calculations: 1) The lobed structure is not due to interference from multiple sources (before and after the singularity), as this mechanism is not included in the model. This confirms that the lobed structure in harmonic vortices results from STCs. 2) For high harmonics in the far-field, the chirp remains independent of harmonic order, $\beta(t) = nct/w_\xi w_x k_0$, with $n$ a constant factor related to the far-field integral. Additionally, as expected, the number of minima between lobes corresponds to the harmonic TC, $l_q = ql$. This confirms that the intensity distribution and spatial chirp can be understood through the FT of the fundamental field.  Furthermore, although no analytical solution exists for the far-field integral of a fundamental PSTOV beam (to our knowledge), the conclusions from the FT of the STOV beam extend to the PSTOV beam, reinforcing the consistency of our approach.  

\textit{Discussion}- Spatiotemporal couplings play a fundamental role in the HHG process when driven by various STOV beams. By analyzing the FT of both STOV and PSTOV beams, we demonstrated that the distinct intensity distributions and spatial chirp observed in these beams result from the presence of STCs in the fundamental beam. Moreover, the calculated intensity distributions clearly show that the TC of a given harmonic scales with the harmonic order and can be determined by counting the minima between lobes. As the TC value increases, the number of minima also increases, a characteristic inherent to STOV beams. Based on these findings, we explained the observed harmonic structure, concluding that the maxima and minima in the intensity distributions arise from TC conservation, which can be mathematically described using the FT. We further concluded that the unique chirped, lobed structure displayed by harmonic vortices is not due to interference between different emission regions of the STOV (before and after the beam's singular region). Additionally, by analyzing numerical results from the QO and TSM approaches, we reconciled differences between the two models, demonstrating that the STCs inherent to the STOV dictate the correct physical interpretation. This work paves the way for new opportunities to investigate and manipulate spatiotemporal vortex beams by leveraging their STCs.

\textit{Acknowledgments}: The present work is supported by the National Key Research and Development Program of China (Grant No.~2023YFA1407100), Guangdong Province Science and Technology Major Project (Future functional materials under extreme conditions - 2021B0301030005) and the Guangdong Natural Science Foundation (General Program project No.~2023A1515010871).

\bibliography{apssamp}

\clearpage

\section*{End Matter}

\subsection{Quantum orbit model}

In the following we will revisit the QO approach to the SFA as presented in Ref.~\cite{PKU}. Within the framework of the QO-SFA, the ionization process, $A_{\text{ion}}(x,t_i)$, is described by:
\begin{eqnarray}
A_{\mathrm{ion}}(x,t_i) &=& C_{n^*,l}\left( \frac{3}{2\pi} \right)E\left( \frac{2(2I_p)^{3/2}}{ E(x,t)} \right)^{(2/ (2I_p)^{1/2})-2}\nonumber \\ 
&\times& \exp \left(-\frac{2(2I_p)^{3/2}}{3E(x,t)} \right)
\end{eqnarray}
with $I_p$, the ionization potential of the electronic ground state and $C_{n^*,l} $ a constant, that depends on the ground state quantum numbers. Here we consider a hydrogen atom with $I_p=0.5$ a.u. The electron propagation in the continuum is described by the term: 
\begin{eqnarray}
A_{\mathrm{prop}}(x,t_i,t_r) = \frac{(2\pi)^{3/2}}{(i(t_r-t_i))^{3/2}}\exp\left( -iS(t_r,t_i) \right).
\end{eqnarray}
Here, we have omitted the dependence on the stationary momentum, as the saddle-point approximation allows us to use $\frac{1}{2}(p_s + A(t_r))^2 + I_p = N\omega$. Similarly, the recombination amplitude can be expressed as \cite{QO}:
\begin{eqnarray}
A_{\mathrm{rec}}(x,t_r) &=& \left( \frac{2\pi}{iS''_{t_r,t_i}} \right) \exp\left( iS(t_r,t'_r) + iN\omega t_r\right)\nonumber \\ 
&\times& d^*(p_s+A(t_r)),
\end{eqnarray}
where $d^*(p_s + A(t_r))$ represents the dipole transition matrix. The recombination amplitude, $A_{\text{rec}}(x, t_r)$, is challenging to calculate because the phase depends on the classical action, $S(t_r, t'_r)$. However, by employing the quantitative re-scattering (QRS) theory presented in Ref.~\cite{QO}, we can assume that $A_{\text{rec}}(x, t_r)$ is independent of the saddle-point momentum. Consequently, we can write:
\begin{equation}
A_{\mathrm{rec}}(x,t_r) = \left( \frac{2\pi}{iS''_{t_r,t_i}} \right) \exp\left(iN\omega t_r\right)d^*(p_s+A(t_r)). 
\end{equation}
Once the contributions of the different amplitudes are known, the dipole response can be calculated as follows \cite{PKU}:
\begin{equation}
    D(x,\omega,m) \propto A_{\mathrm{rec}}(x,t_m^{(r)}) A_{\mathrm{prop}}(x,t_m^{(i)}\rightarrow t_m^{(r)}) A_{\mathrm{ion}}(x,t_m^{(i)}),
\end{equation}
where the ionization and recombination times are denoted by $t_i$ and $t_r$, respectively and $m$ denotes their associated quantum orbit. The final expression of the dipole response is calculated by integrating the complex phase of each amplitude \cite{PKU}. As described in Refs.~\cite{QO} and \cite{PKU}, in the cutoff region, the most relevant trajectories are those originating near the field's maximum and are used to calculate the harmonic response. This simplification implies that the propagation term remains nearly identical across all harmonics. Furthermore, the dipole transition matrix can be approximated, for example, by a Gaussian form \cite{SFA,PKU} or can be used as the exact solution of the Volkov wave projected on the ground state of the hydrogen atom. Here, we tested both possibilities without significative differences in the results. It is important to note that in this model, the coupling between the atomic system and the spatial part of the vortex is neglected; The model uses the specific value of the spatiotemporal field, in matrix form, to calculate the different matrix elements. This means, the different matrix elements are now evaluated in a 2D field, instead of the normal 1D temporal field used in the SFA. This approach enables the direct calculation of the HHG spectrum by summing the contributions from each quantum orbit, $m$, to the total HHG:
\begin{equation}
    P(x,\omega)\propto \omega^4 |\sum_m D(x,\omega,m)|^2. \label{SFA-STOV}
\end{equation}
In addition, the model considers only harmonics in the cutoff region, as multi-recollision processes— which cannot be described by the simple man's model approximation—can be neglected~\cite{QO}. It is also important to remark that even when a full integration is performed for the time variable (saddle point only for the dynamical momentum), no significant differences in the results where observed. More importantly, since the QO incorporates most of the ingredients of the full SFA, the dipole phase effects are automatically incorporated in the results. However, the full propagation effects cannot be incorporated in the model and need to be calculated using, for example, the thin-slab model (TSM) approach. 

\subsection{Thin-slab model}
The thin-slab model (TSM)\cite{AttoVortex}, makes it possible to calculate the macroscopic characteristics of the HHG process driven by structured light. The TSM model can be summarized as follows: We start with the fundamental field, which in this case is a STOV beam described by:  
\begin{eqnarray}
	E_{STOV}(x,y,z,t) &=& A(x,y,z,t) \exp{\left( - i l \phi \right)} \nonumber \\ 
	&\times& \exp{(i k z - \omega_0 t)}. 
\end{eqnarray}
This STOV beam, is used to calculate the harmonic in the near-field by assuming that its main properties are imprinted in the high-order harmonics. In addition, the harmonic amplitude in the near-field is given by the fundamental field amplitude to a power that corresponds to the harmonic scaling-law, $p$. The phase of the harmonic is given by the fundamental field phase scaled with the harmonic order, $q$: 
\begin{eqnarray}
	E^{near}_{STOV}(x,y,z,t) &=& \left(A(x,y,z,t)\right)^p \exp{\left( - i q l \phi \right)} \nonumber \\ 
	&\times& \exp{(-i qk z -i q \omega_0 t)}. 
\end{eqnarray}
For calculating the harmonics in the far-field, it is necessary to solve the far-field Fraunhofer integral, which is given by:
\begin{eqnarray}
   \widetilde{E}^{far}_{STOV}(\beta_x,t)&\propto& A_0\int dx \left(A(x,y,z,t)\right)^p \exp{\left( - i q l \phi \right)}\nonumber \\
    &\times& \exp{\left(-i q \omega_0 t \right)}\exp{(-ik_q \beta_x x)},
\end{eqnarray}
with $A_0 = qk/(2\pi iz)\exp{\left( iqkx^2 / (2z)\right)}$. Notice that to compare the results with the near-field, we have used the integration only over $x$. To our knowledge, the integral has not analytical solution. Consequently, the harmonic vortices are calculated by solving numerically the integral for the far-field $\widetilde{E}^{far}_{STOV}(\beta_x,t)$. The integral can be solved either by using the amplitude term, $Amp$, for a STOV or PSTOV beams. 

\subsection{Fourier transform of the harmonic field}

\begin{figure}[h!]
\centering 
\includegraphics[width=1.0\linewidth]{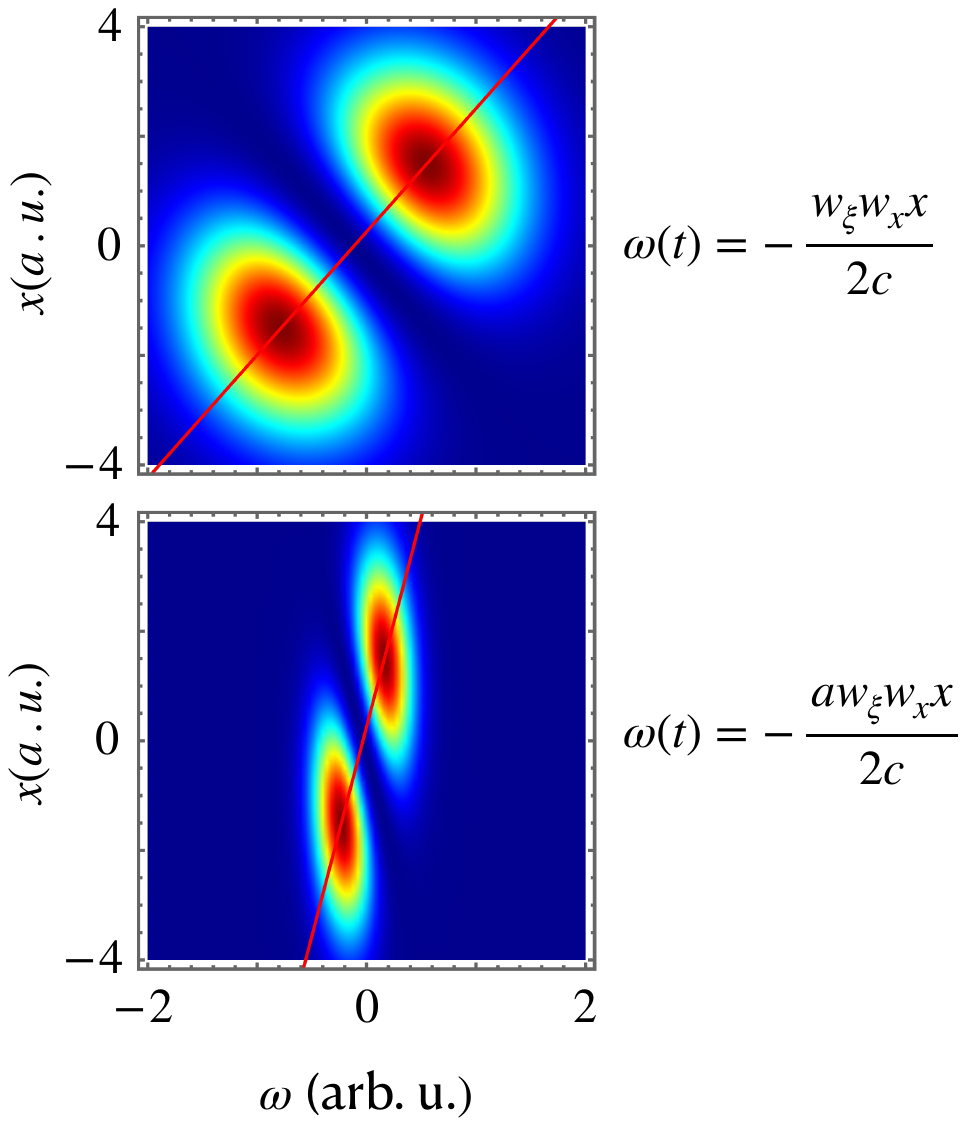}
\caption{Fourier transform of the fundamental field. Here, we present the different chirp functions obtained from the FT without (upper panel) and with (lower panel) the constant value $a$ in the exponent. Note that we centered the intensity distribution around zero to avoid confusion between the central value and the harmonic order. } 
\label{FF3}
\end{figure}

The analysis of fundamental field and its FT (Eqs.~(\ref{stbeam}) and~(\ref{FT_STOV})) let us to the following conclusion: Any integer constant, $a$, multiplying the exponential factor $\exp(-i a \omega t)$ will be reflected in the spatiospectral chirp function by increasing its slope. The constant $c$ is used here instead of the harmonic order. This means that since in the QO approach it is not possible to calculate the individual harmonics and we calculate the full harmonic spectra, without accounting for the individual harmonic order, the QO approach overestimates the chirp function. An example of the analysis is presented in the Fig.~\ref{FF3}. For the upper panel we used $a=1$ and for the lower panel we used $a=17$. This clearly shows that the FT dictates not only the intensity distribution of the different fundamental and harmonic fields, but also, the chirp function. 

\end{document}